\documentstyle[12pt]{article}   

\large
\newcommand{\tiN}{\raisebox{-6.5pt}{$\displaystyle
\stackrel{\displaystyle N}{\sim}$}}

\newcommand{\beq}{\begin{equation}}
\newcommand{\eeq}{\end{equation}}

\makeatletter
\@addtoreset{equation}{section}
\makeatother

\newtheorem{Definition}{Definition}[section]

\def\be{\begin{equation}}
\def\ee{\end{equation}}
\def\ba{\begin{eqnarray}}
\def\ea{\end{eqnarray}}

\def\A{{\cal A}}

\def\ag{{{\cal A}/{\cal G}}}

\def\agb{{\overline {{\cal A}/{\cal G}}}}

\def\Ab{{\overline \A}}

\def\Comp{{\mathchoice
{\setbox0=\hbox{$\displaystyle\rm C$}\hbox{\hbox to0pt
{\kern0.4\wd0\vrule height0.9\ht0\hss}\box0}}
{\setbox0=\hbox{$\textstyle\rm C$}\hbox{\hbox to0pt
{\kern0.4\wd0\vrule height0.9\ht0\hss}\box0}}
{\setbox0=\hbox{$\scriptstyle\rm C$}\hbox{\hbox to0pt
{\kern0.4\wd0\vrule height0.9\ht0\hss}\box0}}
{\setbox0=\hbox{$\scriptscriptstyle\rm C$}\hbox{\hbox to0pt
{\kern0.4\wd0\vrule height0.9\ht0\hss}\box0}}}}
\def\Co{{\mathchoice
{\setbox0=\hbox{$\displaystyle\rm C$}\hbox{\hbox to0pt
{\kern0.4\wd0\vrule height0.9\ht0\hss}\box0}}
{\setbox0=\hbox{$\textstyle\rm C$}\hbox{\hbox to0pt
{\kern0.4\wd0\vrule height0.9\ht0\hss}\box0}}
{\setbox0=\hbox{$\scriptstyle\rm C$}\hbox{\hbox to0pt
{\kern0.4\wd0\vrule height0.9\ht0\hss}\box0}}
{\setbox0=\hbox{$\scriptscriptstyle\rm C$}\hbox{\hbox to0pt
{\kern0.4\wd0\vrule height0.9\ht0\hss}\box0}}}}
\def\Rl{{\mathchoice
{\setbox0=\hbox{$\displaystyle\rm R$}\hbox{\hbox to0pt
{\kern0.4\wd0\vrule height0.9\ht0\hss}\box0}}
{\setbox0=\hbox{$\textstyle\rm R$}\hbox{\hbox to0pt
{\kern0.4\wd0\vrule height0.9\ht0\hss}\box0}}
{\setbox0=\hbox{$\scriptstyle\rm R$}\hbox{\hbox to0pt
{\kern0.4\wd0\vrule height0.9\ht0\hss}\box0}}
{\setbox0=\hbox{$\scriptscriptstyle\rm R$}\hbox{\hbox to0pt
{\kern0.4\wd0\vrule height0.9\ht0\hss}\box0}}}}

\title{QSD VI :\\ 
Quantum Poincar\'e Algebra and a Quantum Positivity 
of Energy Theorem for Canonical Quantum Gravity} 
\author{T. Thiemann\thanks{thiemann@math.harvard.edu}
\thanks{New Address : Albert-Einstein-Institut,
Max-Planck-Institut f\"ur Gravitationsphysik, Schlaatzweg 1, 14473 Potsdam, 
Germany, Internet : thiemann@aei-potsdam.mpg.de} \\
       Physics Department, Harvard University, \\
       Cambridge, MA 02138, USA}
\date{{\small \today\\ Preprint HUTMP-96/B-356}}

\begin{document}

\maketitle

\begin{abstract}
We quantize the generators of the little subgroup of the asymptotic 
Poincar\'e group of 
Lorentzian four-dimensional canonical quantum gravity in the continuum.

In particular, the resulting ADM energy operator is densely defined on an 
appropriate Hilbert space, symmetric and essentially self-adjoint.

Moreover, we prove a quantum analogue of the classical positivity 
of energy theorem due to Schoen and Yau. The proof uses a certain technical
restriction on the space of states at spatial infinity which is suggested
to us given the asymptotically flat structure available. The theorem 
demonstrates 
that several of the speculations regarding the stability of the 
theory, recently spelled out by Smolin, are false once a quantum version
of the pre-assumptions underlying the classical positivity of energy 
theorem is imposed in the quantum theory as well.

The quantum symmetry algebra corresponding to the generators of the 
little group faithfully represents the classical algebra.
\end{abstract}

\section{Introduction}

Following the canonical approach to a quantum theory of gravity, a
(Dirac) observable is by definition a self-adjoint operator on the full 
Hilbert space (not only on the Hilbert space induced from the full Hilbert
space by restricting to the space of solutions of the quantum constraints)
which weakly commutes with the constraint operators. Equivalently, an
observable leaves the physical Hilbert space of solutions to the quantum
constraints invariant.

There are no Dirac observables known in neither classical nor quantum
gravity, except for the asymptotically flat case where it is well-known that
the Poincar\'e generators at spatial infinity form a closed Dirac observable
algebra.

In the present paper we address the question of how to quantize these
generators. We work with the real version \cite{Barbero} of the
originally complex connection formulation of general relativity \cite{0}.
The associated real connection representation has been made a solid
foundation of the quantum theory in the series of papers 
\cite{1,2,3,4,5,6,7}. This rigorous mathematical framework which is based 
on the earlier pioneering work on loop variables, for instance 
\cite{Gambini,RS} for gauge theories and gravity respectively (the latter
of which was designed for the complex variables and thus was lacking an 
appropriate inner product), equips us with the tools necessary to ask the 
question of 
whether the various operators constructed are densely defined, symmetric,
self-adjoint, diffeomorphism invariant and so forth.

But even though this kinematical framework is available, it is still quite
surprising that something like an ADM energy operator can actually be
constructed. The reason for this is that in the representation under
consideration the ADM energy function turns out ot be a rather
non-polynomial function of the canonical momenta and thus it is far from
clear how to define it. It turns out that the same technique that 
enabled one to define the Wheeler-DeWitt operator 
for 3+1 Lorentzian gravity \cite{8,9,10}, the Wheeler-DeWitt operator for 
2+1 Euclidean gravity \cite{10a}, length operators \cite{11}, and matter 
Hamiltonians when coupled to 
gravity \cite{12} can be employed to define Poincar\'e quantum operators.\\
\\
The plan of the present paper is as follows :

In section 2 we review the necessary mathematical background from
\cite{1,2,3,4,5,6,7,15}.

In section 3 we regularize the ADM energy operator. There are at least
two natural orderings, one in which the operator becomes densely defined 
on the full physical Hilbert space as defined in \cite{15} and one in which 
it is not. Nevertheless the latter
operator should be the physically relevant one because when restricting the 
Hilbert space to a subspace which is suggested to us given the 
asymptotically flat structure available, then this operator turns out to be 
positive-semidefinite by 
inspection and essentially self-adjoint on the physical Hilbert space.
This result reveals that several of the speculations spelled out in 
\cite{Smolin} and which were based on several unproved assumptions 
were pre-mature : after taking the quantum dynamics of the theory and the 
quantum asymptotic and regularity conditions on the Hilbert space 
appropriately into account, the quantum positivity of energy theorem is 
{\em not} violated.\\ 
It should be said from the outset, however, that 
the ``quantum positivity of energy theorem" that we provide rests, 
besides on a quite particular regularization procedure of the ADM energy 
operator 
which exploits the fall-off behaviour of the fields at spatial infinity 
quite crucially, on one additional technical assumption (the {\em tangle 
assumption}, see below) whose physical significance is unclear.
Although it can be motivated also given the structure available at 
spatial infinity, it should be stressed that without this assumption 
the positivity theorem would not hold.

In section 4 we naturally extend the unitary representation of the 
diffeomorphism group to include asymptotic translations and rotations 
and compute the symmetry algebra between time translations and the 
spatial Euclidean group, that is, we verify the algebra of the little 
group of the Poincar\'e group. We find no anomaly. As is well-known, the
little group suffices 
to induce the unitary irreducible representations of the Poincar\'e
group. In this paper we do not, however, address the more difficult problem
of how to define a boost quantum operator.

\section{Preliminaries}

We begin with a compact review of the relevant notions from \cite{7,15}.
The interested reader is urged to consult these papers and references 
therein.\\ 
\\ 
We assume that spacetime is of the form $M=\Rl\times \Sigma$ where 
$\Sigma$ has an asymptotically flat topology, that is, there is 
a compact set $B\subset\Sigma$ such that $\Sigma-B$ is homeomorphic with
of $\Rl^3$ with a compact ball cut out. We also assume that $\partial\Sigma$
is homeomorphic with the 2-sphere. The case of more than one component of
$\partial\Sigma$ (e.g. several asymptotic ends or horizons etc.) can be 
treated in a similar way.\\
Denote by $a,b,c,..$ spatial tensor indices and by $i,j,k,..$ $su(2)$ 
indices. The gravitational phase space is described by a canonical pair 
$(A_a^i,E^a_i/\kappa)$ where $A_a^i$ is an $SU(2)$ connection on the 
hypersurface
$\Sigma$, $E^a_i$ is an $\mbox{ad}(SU(2))$ transforming vector density
and $\kappa$ is the gravitational coupling constant.
This means that the symplectic structure is given by
$\{A_a^i(x),E^b_j(y)\}=\kappa\delta(x,y)\delta^b_a\delta^i_j$. The 
relation with the co-triad $e_a^i$, the extrinsic curvature $K_{ab}$
and the three-metric $q_{ab}=e^i_a e^i_b$ is 
$E^a_i=\frac{1}{2}\epsilon^{abc}\epsilon_{ijk}e^j_b e^k_c$ and 
$A_a^i=\Gamma_a^i+\mbox{sgn}(\det((e_c^j)))K_{ab} e^b_j$ where
$\epsilon^{abc}$ is the metric independent, completely skew tensor density
of weight one, $\Gamma_a^i$ is the spin-connection of $e_a^i$ and $e^a_i$
is the inverse of the matrix $e_a^i$. 

By $\gamma$ we will denote in the sequel a closed, piecewise analytic graph
embedded into a d-dimensional smooth manifold $\Sigma$ (the case of interest
in general relativity is $d=3$).
The set of its edges will be denoted $E(\gamma)$ and the set of its 
vertices $V(\gamma)$. By suitably subdividing edges into two halves we 
can assume that all of them are outgoing from a vertex (the remaining 
endpoint of the so divided edges is not a vertex of the graph because 
it is a point of analyticity). Let $A$ be a $G$-connection for a compact 
gauge group
$G$ (the case of interest in general relativity is $G=SU(2)$). We will denote
by $h_e(A)$ the holonomy of $A$ along the edge $e$. Let $\pi_j$ be the 
(once and for all fixed representant of the equivalence class of the 
set of) $j$-th irreducible representations of $G$ (in general relativity
$j$ is just a spin quantum number) and label each edge $e$ of $\gamma$ 
with a label $j_e$. Let $v$ be an 
$n$-valent vertex of $\gamma$ and let $e_1,..,e_n$ be the edges incident 
at $v$. Consider the decomposition of the tensor product 
$\otimes_{k=1}^n\pi_{j_{e_k}}$ into irreducibles and denote by 
$\pi_{c_v}(1)$ the
linearly independent projectors onto the irreducible representations $c_v$ 
that appear. 
\begin{Definition} \label{def0}
An extended spin-network state is defined by
\be \label{1}
T_{\gamma,\vec{j},\vec{c}}(A):=
\mbox{tr}(\otimes_{v\in V(\gamma)}
[\pi_{c_v}(1)\cdot\otimes_{e\in E(\gamma),v\in e}\pi_{j_e}(h_e(A))])
\ee
where $\vec{j}=\{j_e\}_{e\in E(\gamma)},\;\vec{c}=\{c_v\}_{v\in 
V(\gamma)}$. In what follows we will use a compound label
$I\equiv(\gamma(I),\vec{j}(I),\vec{c}(I))$. An ordinary spin-network
state is an extended one with all vertex projectors corresponding
to singlets.
\end{Definition}
Thus, a spin-network state is a particular function of 
smooth connections restricted to a graph with definite transformation 
properties under gauge transformations at the vertices. Their importance is 
that they form
an orthonormal basis for a Hilbert space ${\cal H}\equiv{\cal H}_{aux}$, 
called the auxiliary Hilbert space. Orthonormality means that
\be \label{2}
<T_{\gamma,\vec{j},\vec{c}},T_{\gamma',\vec{j}',\vec{c}'}>
\equiv
<T_{\gamma,\vec{j},\vec{c}},T_{\gamma',\vec{j}',\vec{c}'}>_{aux}
=\delta_{\gamma\gamma'}\delta_{\vec{j},\vec{j}'}
\delta_{\vec{c},\vec{c}'}\;.
\ee
Another way to describe $\cal H$ is by displaying it as a space of
square integrable functions $L_2(\agb,d\mu_0)$. Here $\agb$ is a space 
of distributional connections modulo gauge transformations, typically 
non-smooth and $\mu_0$ is a 
rigorously defined, $\sigma$-additive, diffeomorphism invariant probability
measure on $\agb$. The space $\agb$ is the maximal extension of 
the space $\ag$ of smooth connections such that (the Gel'fand 
transform of) spin-network functions 
are still continuous. The inner product can be extended, with the same
orthonormality relations, to any smooth (rather than analytic) graph with a 
finite number of edges and to non-gauge-invariant functions. It is only  
the latter description of $\cal H$ which enables one to verify that 
the inner product $<.,.>$ is the unique one that incorporates the 
correct reality conditions that $A,E$ are in fact real valued. The
inner product (\ref{2}) was postulated earlier (see remarks in \cite{RDP})
for the {\em complex} connection formulation. But
it was not until the construction of the Ashtekar-Lewandowski measure $\mu_0$
that one could show that this inner product is actually the correct one 
for the real connection formulation only.

We will denote by $\Phi$ the finite linear combinations of spin-network 
functions and call it the space of cylindrical functions. A function 
$f_\gamma$ is said to be cylindrical with respect to a graph 
$\gamma$ whenever it is a linear combination of spin-network functions on 
that graph such that $\pi_{j_e}$ is not the trivial representation for no
$e\in E(\gamma)$. 
The space $\Phi$ can be equipped with one of the standard nuclear topologies
induced from $G^n$ because on each graph $\gamma$ every cylindrical 
function $f_\gamma$ becomes a function $f_n$ on $G^n$ where $n$ is the 
number of 
edges $e$ of $\gamma$ through the simple relation $f_\gamma(A)=
f_n(h_{e_1}(A),..,h_{e_n}(A))$. This 
turns it into a topological vector space. By $\Phi'$ we mean the topological
dual of $\Phi$, that is, the bounded linear functionals on $\Phi$. General
theorems on nuclear spaces show that the inclusion $\Phi\subset{\cal 
H}\subset\Phi'$ (Gel'fand triple) holds.

So far we have dealt with solutions to the Gauss constraint only, that is,
we have explicitly solved it by dealing with gauge invariant functions only.
We now turn to the solutions to the diffeomorphism constraint (we follow 
\cite{15}).\\ 
Roughly speaking one constructs a certain subspace $\Phi_{Diff}$ of 
$\Phi'$ by ``averaging spin-network states
over the diffeomorphism group" by following the subsequent recipe :\\
Take a spin-network state $T_I$ and consider its orbit $\{T_I\}$ under the 
diffeomorphism group. Here we mean orbit under asymptotically identity 
diffeomorphisms only ! Then construct the distribution
\be \label{3}
[T_I]:=\sum_{T\in\{T_I\}} T
\ee
which can be explicitly shown to be an 
element of $\Phi'$. Any other vector is averaged by 
first decomposing it into spin-network states and then averaging those 
spin-network states separately. Certain technical difficulties having to 
do with superselection rules and graph symmetries \cite{7} were removed in 
\cite{15}.

An inner product on the space of the so constructed states is given by\\
\be \label{4}
<[f],[g]>_{Diff}:=[f](g)
\ee
where the brackets stand for the averaging 
process and the right hand side means evaluation of a distribution on a 
test function. The completion of $\Phi_{Diff}$ with respect to 
$<.,.>_{Diff}$ is denoted ${\cal H}_{Diff}$.

Finally, the Hamiltonian constraint is solved as follows \cite{10} :
One can explicitly write down an algorithm of how to construct the most
general solution. It turns out that 
one can construct ``basic" solutions $s_\mu\in\Phi'$ which are
mutually orthonormal with respect to $<.,.>_{Diff}$ (in a generalized sense)
and diffeomorphism invariant.
The span of these solutions is equipped 
with the natural orthonormal basis $s_\mu$ (in the generalized sense).
One now defines a ``projector" 
\be \label{5}
\hat{\eta}f:=[[f]]:=\sum_\mu s_\mu <s_\mu,[f]>_{Diff}
\ee
for each $f\in\Phi$ and so obtains a 
subspace $\Phi_{Ham}\subset\Phi'$. 
The physical inner product \cite{15} is defined by 
\be \label{6}
<[[f]],[[g]]>_{phys}:=[[f]]([g])\;.
\ee
Finally, the physical Hilbert space is just the completion of 
$\Phi_{Ham}$ with respect to $<.,.>_{Ham}$.

\section{Regularization of the ADM Hamiltonian}

There are many ways to write the ADM-Hamiltonian 
which are all classically weakly identical. We are going to choose a form 
which is a pure surface integral and which depends only on $E^a_i$ 
because in this case the associated operator will be almost diagonal in a 
spin-network basis so that we can claim that spin-network states really
do provide a non-linear Fock representation for quantum general relativity
as announced in \cite{8,9}.

The appropriate form of the classical symmetry generators was derived in 
\cite{13,13a}.
Although that paper was written for the {\em complex} Ashtekar variables,
all results can be taken over by carefully removing factors of $i$ at 
various places. We find for the surface part of the Hamiltonian (expression 
(4.31) in \cite{13}, we 
use that $\tiN=N/\sqrt{\det(q)},\;D_a\tiN=(D_a N)/\sqrt{\det(q)}$ where
$N$ is the scalar lapse function)
\be \label{20}
E(N)=-\frac{2}{\kappa}\int_{\partial\Sigma} dS_a\frac{N}{\sqrt{\det(q)}}
E^a_i\partial_b E^b_i\;.
\ee
It is easy, instructive and for the sign of the ADM energy crucial to see 
that (\ref{20}) really equals 
the classical expression $+\frac{1}{\kappa}\int_{\partial\Sigma}dS_a 
(q_{ab,b}-q_{bb,a})$ due to ADM :
Using that $E^a_i=\frac{1}{2}\epsilon^{abc}\epsilon_{ijk} e_a^j e_b^k$
we have the chain of identities
\ba \label{7}
&&-\frac{2}{\sqrt{\det(q)}}E^a_i\partial_b E^b_i
=-\mbox{sgn}(\det(e))e^a_i\epsilon^{bcd}\epsilon_{ijk}[e_c^j e_d^k]_{,b}
\nonumber\\
&=&-2\mbox{sgn}(\det(e))e^a_i\epsilon^{bcd}\epsilon_{ijk}e_c^j e_{d,b}^k
\nonumber\\
&=& -2\mbox{sgn}(\det(e))q^{af}\epsilon^{bcd}\epsilon_{ijk}e_f^i e_c^j 
e_{d,b}^k
=-2q^{af}\epsilon^{bcd}\sqrt{\det(q)} \epsilon_{fce}e^e_k e_{d,b}^k
\nonumber\\
&=&-4q^{ac}\delta^b_{[c}\delta^d_{e]}\sqrt{\det(q)} e^e_i e_{d,b}^i
\nonumber\\
&=& 4\sqrt{\det(q)} q^{ac}  q^{ed} e_d^i e_{[c,e]}^i
=2\sqrt{\det(q)} q^{ac}  q^{bd} e_d^i[e_{c,b}^i-e_{b,c}^i]
\nonumber\\
&=& \sqrt{\det(q)} q^{ac}  q^{bd} 
[2e_{(d}^i e_{c),b}^i+2e_{[d}^i e_{c],b}^i
-2e_{(d}^i e_{b),c}^i-2e_{[d}^i e_{b],c}^i]
\nonumber\\
&=& \sqrt{\det(q)} q^{ac}  q^{bd} 
[(q_{cd,b}-q_{bd,c})
+2e_{[d}^i e_{c],b}^i]\;
\ea
Now we expand $e_a^i(x)=\delta_a^i+\frac{f_a^i(x/r)}{r}+o(1/r^2)$ where
$r^2=\delta_{ab}x^a x^b$ defines the asymptotic Cartesian frame. The function
$f_a^i(x/r)$ only depends on the angular coordinates of the asymptotic 
sphere and is related to the analogous expansion 
$q_{ab}(x)=\delta_{ab}+\frac{f_{ab}(x/r)}{r}+o(1/r^2)$ by 
$f_{ab}\delta^b_i=f_a^i$. Consider now the 
remainder in the last line of (\ref{7}). Its $o(1/r^2)$ part vanishes 
because $f_{[ab]}=0$ and this concludes the proof.

In the sequel we focus on the energy functional $E_{ADM}=E(N=1)$. We will 
quantize it in two different ways correspoinding to two quite different 
factor orderings. Each of the orderings has certain advantages and 
certain disadvantages which we will point out.

\subsection{Ordering I : No state space restriction}

In this subsection we will derive a form of the operator which is densely 
defined on the whole Hilbert space $\cal H$ (and extends to the spaces 
${\cal H}_{Diff},\;{\cal H}_{phys}$ defined above) without imposing any further
restriction that corresponds to asymptotic flatness.

Using again that
$E^a_i=\frac{1}{2}\epsilon_{ijk}\epsilon^{abc}e^j_b e^k_c$ we can write 
it as
\be \label{21}
E_{ADM}=\lim_{S\to \partial\Sigma} E_{ADM}(S) 
\mbox{ where } E_{ADM}(S)=-\frac{2}{\kappa}\int_S
\frac{1}{\sqrt{\det(q)}} \epsilon^{ijk}e^j\wedge e^k \partial_b E^b_i
\ee
and $S$ is a closed 2-surface which is topologically a sphere.
The idea is to point-split expression (\ref{21}) and to use that
$$
[\mbox{sgn}(\det(e))e_a^i](x)=\frac{1}{2\kappa}\{A_a^i(x),V(x,\epsilon)\}
$$
where $V(x,\epsilon)=\int_\Sigma d^3y \chi_\epsilon(x,y)\sqrt{\det(q)}(y)$
and $\chi_\epsilon$ is the (smoothed out) characteristic function of a 
box of coordinate volume $\epsilon^3$. Since 
$$
\lim_{\epsilon\to 0}\frac{\chi_\epsilon(x,y)}{\epsilon^3}=\delta(x,y)
\mbox{ so that } \lim_{\epsilon\to 0}\frac{V(x,\epsilon)}{\epsilon^3}
=\sqrt{\det(q)}(x)
$$
we have
\ba \label{22}
&&E_{ADM}(S)\nonumber\\
&=&\lim_{\epsilon\to 0}
-\frac{2}{\kappa}\int_S 
\frac{1}{\epsilon^3\sqrt{\det(q)}(x)}
\epsilon^{ijk}e^j(x)\wedge e^k(x)\int_\Sigma d^3y
\chi_\epsilon(x,y)(\partial_b E^b_i)(y)
\nonumber\\
&=& \lim_{\epsilon\to 0}-\frac{2}{\kappa}\int_S 
\frac{\epsilon^{ijk}}{V(x,\epsilon)}
e^j(x)\wedge e^k(x)\int_\Sigma d^3y \chi_\epsilon(x,y)(\partial_b E^b_i)(y)
\nonumber\\
&=& \lim_{\epsilon\to 0}
-\frac{1}{2\kappa^3}\int_S 
\frac{\epsilon^{ijk}}{V(x,\epsilon)}
\{A^j(x),V(x,\epsilon)\}\wedge  \{A^k(x),V(x,\epsilon)\}
\int_\Sigma d^3y \chi_\epsilon(x,y) (\partial_b E^b_i)(y) \nonumber\\
&=& \lim_{\epsilon\to 0}
-\frac{2}{\kappa^3}\int_S \epsilon^{ijk} 
\{A^j(x),\sqrt{V(x,\epsilon)}\}\wedge \{A^k(x),\sqrt{V(x,\epsilon)}\}
\int_\Sigma d^3y \chi_\epsilon(x,y)
(\partial_b E^b_i)(y) \nonumber\\
&=& \lim_{\epsilon\to 0}
\frac{4}{\kappa^3}\int
\mbox{tr}(\{A(x),\sqrt{V(x,\epsilon)}\}\wedge 
\{A(x),\sqrt{V(x,\epsilon)}\} \int_\Sigma d^3y \chi_\epsilon(x,y)
(\partial_b E^b)(y)) \nonumber\\
&=& -\lim_{\epsilon\to 0}
\frac{4}{\kappa^3}\int_S 
\mbox{tr}(\{A(x),\sqrt{V(x,\epsilon)}\}\wedge 
\{A(x),\sqrt{V(x,\epsilon)}\} \int_\Sigma d^3y 
[\partial_{y^b}\chi_\epsilon(x,y)] E^b(y)) \nonumber\\
&=& \lim_{\epsilon\to 0} E^\epsilon_{ADM}(S)
\ea
where in the second before the last step we have taken a trace with 
respect to generators $\tau_i$ of $su(2)$ obeying 
$[\tau_i,\tau_j]=\epsilon_{ijk}\tau_k$ and in 
the last step we have performed an integration by parts
(the boundary term at $\partial\Sigma$ does not contribute for finite $S$
and $\epsilon$ sufficiently small).
Thus, we absorbed the $1/\sqrt{\det(q)}$ into a square-root within a 
Poisson-bracket and simultanously the singular $1/\epsilon^3$ into a 
volume functional. Classically we could have dropped the $1/\sqrt{\det(q)}$
(although the integrand would then no longer be a density of weight one 
and is strictly speaking not the boundary integral of a variation of the 
Hamiltonian constraint) due to the classical boundary conditions which 
tell us that $\det(q)$ tends to $1$. \\
We now quantize $E^\epsilon_{ADM}(S)$. This consists of two parts : In the 
first we focus on the volume integral in (\ref{22}) and replace $E^a_i$ by
$\hat{E}^a_i=-i\hbar\kappa\delta/\delta A_a^i$. In the second step we 
triangulate $S$ exactly as the hypersurface of 2+1 gravity in \cite{10a}, 
replace the volume 
functional by the volume operator and Poisson brackets by commutators times
$1/(i\hbar)$. 

So let $f_\gamma$ be a function cylindrical with 
respect to a graph $\gamma$. Since we are only interested in the 
limit $S\to\partial\Sigma$ we may assume that \\
1) $\gamma$ lies entirely within the closed ball whose boundary is $S$ but\\
2) $\gamma$ may intersect $S$ at an endpoint of one of its edges and may 
even have edges that lie entirely inside $S$.\\
Furthermore we can label the edges of $\gamma$ in 
such a way that an edge either intersects $S$ transversally (with an 
orientation outgoing from the intersection point with $S$) or lies entirely 
within $S$.\\ 
Coming to the first step we have for the $y$ integral involved in
$E^\epsilon_{ADM}(S)$ :
\ba \label{23}
&& \int_\Sigma d^3y [\partial_a\chi_\epsilon(x,y)]
\hat{E}^a_(y)_i f_\gamma \nonumber\\
&=& -i\hbar\kappa \sum_{e\in E(\gamma)}\int_\Sigma d^3y 
[\partial_a\chi_\epsilon(x,y)]
\int_0^1 dt \dot{e}^a(t) \delta(y,e(t))X^i_e(t) f_\gamma 
\nonumber\\
&=& -i\hbar\kappa \sum_{e\in E(\gamma)}\int_0^1 dt \dot{e}^a(t)
[\partial_{y^a}\chi_\epsilon(x,y)_{y=e(t)}]
X^i_e(t) f_\gamma 
\nonumber\\
&=& -i\hbar\kappa \sum_{e\in E(\gamma)}\int_0^1 dt 
[\frac{d}{dt}\chi_\epsilon(x,e(t))]
X^i_e(t) f_\gamma 
\nonumber\\
&=& -i\hbar\kappa \sum_{e\in E(\gamma)} \lim_{n\to\infty}
\sum_{k=1}^n [\chi_\epsilon(x,e(t_k))-\chi_\epsilon(x,e(t_{k-1}))]
X^i_e(t_{k-1}) f_\gamma 
\ea
where $E(\gamma)$ is the set of edges of $\gamma$, $X^i_e(t):=
[h_e(0,t)\tau_i h_e(t,1)]_{AB} \partial/\partial[h_e(0,1)]_{AB}$
and $0=t_0<t_1<..<t_n=1$ is an arbitrary partition of the interval 
$[0,1]$.\\
It is important for what follows that for each $t,e,i$ $X^i_e(t) f_\gamma$
is still a function cylindrical with respect to $\gamma$.

We now come to the second step. This involves, first of all, a 
triangulation of the {\em two-dimensional} surface $S$ in adaption to the
graph $\gamma$. Besides the prescription explained in detail in \cite{10a}
which deals with the triangulation of $S$ in the neighbourhood of a 
vertex formed by edges of $\gamma$ that lie entirely within $S$ we 
just need to deal with the case that a vertex $v$ of $\gamma$ has also edges
incident at it which lie entirely inside the open ball with boundary $S$ 
except for the one point $v$. In case that there are at least {\em two}
edges $e_1,e_2$ of $\gamma$ incident at $v$ such that $e_1,e_2\subset S$ we can 
still take over the triangulation from \cite{10a}. However, if there is
only {\em one or no} such edge (indeed, since we do not allow for gauge 
transformations at spatial infinity we can allow for open edges that lie 
entirely within or end at $S$ without ruining gauge invariance) we need 
an additional prescription : In case there is only one edge $e\subset S$ 
incident
at $v$, choose an arbitrary edge $e'$ not intersecting $\gamma$ except 
at $v$ such that the tangents of the edges $e,e',e^{\prime\prime}$ are 
positively 
oriented at $v$ where $e^{\prime\prime}$ is any of the edges of $\gamma$ 
incident at $v$ but transversal to $S$.\\ 
In case there is no edge $e\subset S$ incident
at $v$, choose two arbitrary edges $e,e'$ not intersecting $\gamma$ 
except at $v$ such that  
the tangents of the edges $s,e',e^{\prime\prime}$ are positively 
oriented at $v$ where 
$e^{\prime\prime}$ is any of the edges of $\gamma$ incident at $v$ but 
transversal to $S$.
These arbitrary edges will disappear from the stage again at the end of 
the calculation.\\
Given this set-up, at each vertex $v$ of $\gamma$ that lies inside $S$ we 
have now at least to edges $e_1,e_2$ incident at it that are inside $S$ 
and we can 
define the triangles $\Delta$ associated with pairs of edges incident at 
$v$ and inside $S$ exactly as in \cite{10a}. As in $\cite{10a}$ we then 
have two segments $s_1(\Delta),s_2(\Delta)$ for each triangle $\Delta$
which are actually segments of edges of $\gamma$ incident at $v$ and that 
lie inside $S$. Now observe that
\ba \label{24}
&& \epsilon^{ij} h_{s_i(\Delta)}\{h_{s_i(\Delta)}^{-1},V(v,\epsilon)\}
h_{s_j(\Delta)}\{h_{s_j(\Delta)}^{-1},V(v,\epsilon)\}
\nonumber\\
&=&\delta^2\epsilon^{ij}\dot{s}^a_i(0)\dot{s}^b_j(0)
\{A_a(v),\sqrt{V(v,\epsilon)}\}\{A_b(v),\sqrt{V(v,\epsilon)}\}
+o(\delta^2)\nonumber\\
&=& 2\mbox{vol}(\Delta)\epsilon^{ab}
\{A_a(v),\sqrt{V(v,\epsilon)}\}\{A_b(v),\sqrt{V(v,\epsilon)}\}
\ea
where $\delta$ is a small parameter corresponding to the parameter length
of the $s_i(\Delta)$ and $\epsilon^{ab}$ the metric independent totally 
skew tensor density of weight one on $S$. Altogether we therefore conclude 
that the surface integral 
$\int_S \{A(x),\sqrt{V(x,\epsilon)}\}\wedge \{A(x),\sqrt{V(x,\epsilon)}\}$
involved in $E_{ADM}(S)$ can be quantized by
\be \label{25}
-\frac{1}{2\hbar^2} \sum_{v\in V(\gamma)} \frac{4}{E(v)}\sum_{v(\Delta)=v}
\epsilon^{ij} 
h_{s_i(\Delta)}[h_{s_i(\Delta)}^{-1},\sqrt{\hat{V}(v,\epsilon)}]
h_{s_j(\Delta)}[h_{s_j(\Delta)}^{-1},\sqrt{\hat{V}(v,\epsilon)}] f_\gamma\;.
\ee
We do not wish to give a full derivation of (\ref{25}) for which the 
reader should consult \cite{10a}, however, a few remarks are in order
which intuitively explain (\ref{25}) :
\begin{itemize}
\item[1)] First of all as already mentioned, we do not work with a fixed 
triangulation but with a whole finite family of triangulations that 
depend on the 
graph $\gamma$ of the state that we act on. More precisely, a particular 
member of the family of triangulations is defined 
such that for each vertex $v$ of $\gamma$ which lies in $S$ we a) pick one 
pair $e_1,e_2$ of edges of $\gamma$ incident at it which are such that 
$e_2$ is next to $e_1$ to its right while the tangents of $e_1,e_2$ 
enclose an angle less than or equal to $\pi$ (with respect to 
$\delta_{ab}$), b) take proper 
subsegments $s_i(\Delta)\subset e_i$ incident at $v$, c) construct two more
segments $s_{\bar{i}}(t)=2v-s_i(t)$, d) construct four obvious 
triangles from $s_i,s_{\bar{i}}$ which saturate $v$ and e) 
choose a triangulation which embeds those four triangles for each $v$
as basic ones and is otherwise arbitrary only subject to the restriction 
that none of the remaining $\Delta$ has its basepoint on $\gamma$. 
\item[2)] Next, each of the four triangles 
$\Delta_{12},\Delta_{2\bar{1}},\Delta_{\bar{1}\bar{2}},\Delta_{\bar{2}1}$
contributes classically the same to the surface 
integral $\int_S=\sum_\Delta \int_\Delta$ as the triangle $\Delta_{12}$
so that we can classically replace these four terms by
$4\int_{\Delta_{12}}$. This explains the factor of $4$ in (\ref{25}).
\item[3)] Now, since we do not want to distinguish one particular pair of 
edges as 
compared to any other, we average over the choice of pairs which explains 
the factor of $1/E(v)$ where $E(v)=n(v)$ or $E(v)=n(v)-1$ is the 
possible number of pairs depending on whether the angles less than or 
equal to $\pi$ add up to $2\pi$ (with respect to $\delta_{ab}$) or not and 
$n(v)$ is the valence of $v$.
\item[4)] Finally, the fact that we sum over vertices of $\gamma$ only comes 
from the presence of the volume operator which has non-trivial action at 
vertices only. Therefore the contribution of all other triangles which 
define the triangulation drop out (all triangles such that there is no 
edge of $\gamma$ transversal to $S$ at its basepoint since the 
three-dimensional volume operator annihilates co-planar vertices).
\end{itemize}
It should be mentioned that any of the so-defined ``averaged family of 
triangulations of $S$ adapted to $\gamma$" has as classical continuum limit 
the original integral $\int_\Sigma$ !\\
Putting (\ref{23}) and (\ref{25}) together we obtain as the final result
\ba \label{26}
\hat{E}^{\epsilon,n}_{ADM}(S)f_\gamma&=&
\frac{4}{\kappa^3}(-\frac{1}{2\hbar^2}) (-i\hbar\kappa)
\sum_{v\in V(\gamma)} \frac{4}{E(v)}\sum_{v(\Delta)=v}\times\nonumber\\
&\times&\epsilon^{ij} 
\mbox{tr}(h_{s_i(\Delta)}[h_{s_i(\Delta)}^{-1},\sqrt{\hat{V}(v,\epsilon)}]
h_{s_j(\Delta)}[h_{s_j(\Delta)}^{-1},\sqrt{\hat{V}(v,\epsilon)}]
\times\nonumber\\
&\times&\sum_{e\in E(\gamma)} 
\sum_{k=1}^n [\chi_\epsilon(v,e(t_k))-\chi_\epsilon(v,e(t_{k-1}))]
X^i_e(t_{k-1}) f_\gamma \;.
\ea
Now we perform the limit $n\to\infty$ and $\epsilon\to 0$ in reversed 
order\footnote{One can also make $\epsilon$ $y$-dependent 
in (\ref{22}) in a state-dependent way which then leads to a dependence 
$\epsilon=o(1/n)$. If one then takes $n\to\infty $ one gets  
classically back to the original expression for 
the ADM energy. If one takes $n$ large but finite then one arrives 
at the same result as below on the quantum level 
without that an interchange of limits is necessary. Therefore 
the calculations that follow are justified. The state-dependence of the 
regularization drops out in the final expression because before actually
taking the limit $n\to\infty$ the operator will be $n$-independent. The
family of operators so obtained for each state are consistently defined 
as we will see. See \cite{10a,12} for a more detailed explanation.} : 
Keeping $n$ fixed, for small enough $\epsilon$ only the term with
$k=1$ in the sum survives provided that $e(0)=v$. Therefore also the sum
over $k$ and with it the $n$ dependence drops out. Also the operator
$\hat{V}(v,\epsilon)$ actually has a limit as $\epsilon\to 0$ which we call
$\hat{V}_v$ and which is defined by $\lim_{\epsilon\to 0}
\hat{V}(v,\epsilon)f_\gamma=:(\hat{V}_v)_\gamma f_\gamma$ for all $\gamma$. 
That 
the limit exists relies on the fact that either $\gamma$ has $v$ as a vertex
or it does not. In the latter case the limit just vanishes, in the former 
case for sufficiently small $\epsilon$ the $\epsilon$ box around $v$ does 
not include any other vertex of $\gamma$ other than $v$ and so there is only
one contribution $(\hat{V}_v)_\gamma f_\gamma$ which is  constant as 
$\epsilon\to 0$. The family of operators $(\hat{V}_v)_\gamma$ is 
consistently defined because $\hat{V}$ is. For a more explicit formula 
in terms of analytic germs of edges see \cite{10a}.\\
Now, the limit $n\to\infty$ is trivial and the resulting operator derived 
for arbitrary but finite $S$ can be extended to $\partial\Sigma$. The 
result is
\ba \label{27}
\hat{E}_{ADM} f_\gamma&=&
\frac{2i}{\hbar\kappa^2}
\sum_{v\in V(\gamma)} \frac{4}{E(v)}\sum_{v(\Delta)=v}\times\nonumber\\
&\times&\epsilon^{ij} 
\mbox{tr}(h_{s_i(\Delta)}[h_{s_i(\Delta)}^{-1},\sqrt{\hat{V}(v,\epsilon)}]
h_{s_j(\Delta)}[h_{s_j(\Delta)}^{-1},\sqrt{\hat{V}(v,\epsilon)}]
\times\nonumber\\
&\times&\sum_{e\in E(\gamma),e(0)=v} 
X^i_e f_\gamma 
\ea
where $X^i_e(0)=X^i_e=X^i(h_e))$ and $X^i(g)$ is the right invariant 
vector field at $g\in SU(2)$.\\
The virtue of (\ref{27}) is that it displays the ADM energy operator as a
densely defined operator on all of the Hilbert space. Also, the 
dependence on the ``arbitrarily short edges $s_i(\Delta)$ drops out at 
the end of the calculation because of gauge invariance as explained 
in \cite{11}. The disadvantage is that the operator (\ref{27}) is not 
a manifestly positive semi-definite operator. This is, however, not 
surprising because even the classical ADM energy is not a positive 
semi-definite functional on the full phase space of general relativity.
It is only when evaluating it on a) asymptotically flat b) solutions of the 
Einstein equations which c) satisfy an energy condition for allowed matter 
and d) allow for a regular initial data set, that the positive energy 
theorem has been proved \cite{Yau,Witten} and in fact one can easily
produce negative ADM energy when one of these conditions is violated.
As we did not impose any (quantum analogue of) such restrictions we 
cannot expect to find a manifestly non-negative operator. \\
In the next subsection we will derive another quantization of the 
ADM energy which is only densely defined on a subspace of the Hilbert space,
however, the definition of that subspace a quite natural quantum 
translation of the classical condition that there be an asymptotically flat 
regular initial data set. The virtue will be that the ADM operator 
acquires non-negative discrete spectrum on that subspace of the Hilbert 
space thus proving a ``Quantum Positivity of Energy Theorem".

\subsection{Ordering II : Restrictions on the State Space}

In order to make sense of the operator to be defined in this section
we need to give some definition of ``asymptotically flat state". The 
following definition is a first attempt towards a precise notion of 
``quantum asymptotic flatness" to be considerably refined in future 
publications. 
\begin{Definition} \label{def1}
An asymptotically flat state $\Psi$ on the Hilbert space $\cal H$ is 
a distribution in $\Phi'$ satisfying the following conditions :\\
i) $\Psi$ is a normalized solution $\hat{\eta}\xi$ to all quantum 
constraints 
$\hat{U}(\varphi)-1=0,\;\hat{H}(N)=0$ (the diffeomorphism and Hamiltonian 
constraint) of general 
relativity where $\varphi,N$ are arbitrary 
three-diffeomorphisms and lapse functions subject to the 
condition that they be pure gauge, that is, $\varphi(x)\to 
\mbox{id}$ and $N(x)\to 0$ as $x\to \partial \Sigma$ and 
where $\hat{\eta}$ is 
the operator \cite{15} 
that maps 
elements of $\Phi$ to solutions to all constraints.\\
ii) $\Psi=\hat{\eta}\xi$ is asymptotically flat. That is, consider any 
compact region $R$, surface $S$ and loop $\alpha$ 
(which is embedded in the graph underlying the definition of $\xi$)
of $\Sigma-\partial\Sigma$ 
and which are large as compared to Planck volume, area and length 
respectively as measured by the 
Euclidean metric $q^{(0)}_{ab}:=\delta_{ab}$.
Then, as $R,S,\alpha$ tend to $\partial\Sigma$, the quantities
$\Psi([\hat{V}(R)-V_0(R)]\xi),\Psi([\hat{A}(S)-A_0(S)]\xi),
\Psi([\mbox{tr}(h_\alpha)-2]\xi)$ respectively 
are of order $\ell_p^3,\ell_p^2,\ell_p/L(c)$ respectively 
where $V_0(R),A_0(S),L_0(\alpha)$ are the volume, area and length of 
$R,S,\alpha$ as measured by $q^{(0)}_{ab}$.\\
iii) $\Psi$ transfroms according to an unitary, irreducible 
representation of the Poincar\'e group at spatial infinity.\\
iv) $\Psi$ satisfies the dominant energy condition :
Let $\hat{H}_{matter}'(N)$ be the dual of the matter part of the 
Hamiltonian constraint (not Hamiltonian !) and 
$\hat{V}_{matter}'(\vec{N})$ be the dual of the matter part of the  
Diffeomorphism constraint.\\ 
The four-vector $N^\mu=(N,\vec{N})$ is said to 
be a future directed timelike vector in a state $\Psi$ if a) $N>0$ and   
b) there exists $\epsilon>0$ such that $-t^2 
N^2(x)+\Psi[\hat{L}(c(\vec{N},x,t))^2\xi]<0$ for each $0<t<\epsilon$
where $\hat{L}(c)$ is the length operator \cite{11} and $c(\vec{N},x,t)$
is the segment of the integral curve of $\vec{N}$ beginning in $x$ and 
ending after a parameter distance $t$.\\
$\Psi$ is said to satisfy the dominant energy condition provided that
for every four-vector $N^\mu$ which is future directed and timelike for 
$\Psi$ then there is an $\epsilon>0$ such that the four vector 
$\hat{P}^\mu(x,t):=(\hat{H}_{matter}'(N_{x,t})
+\hat{V}_{matter}'(\vec{N}_{x,t}),0)$ is 
either zero or 
future directed and timelike for every $x\in\Sigma$ and $0<t<\epsilon$ in 
the state $\Psi$ where 
$N_{x,t}(y)=\chi_t(x,y)N(y)$ and likewise for $\vec{N}_{x,t}(y)$
($\chi_t(x,y)$ is the characteristic function of a box of coordinate volume
$t^3$ and centre $x$).
In other words,
$\Psi([\hat{H}_{matter}(N_{x,t})+\hat{V}_{matter}(\vec{N}_{x,t})]\xi)>0$ for 
each $x,0<t<\epsilon$. Here we have adopted the convention that the 
signature of the Lorentz metric be $-,+,+,+$.\\
The subspace of $\xi$'s 
in $\cal H$ satisfying these conditions will be called ${\cal H}_{af}$
where ``af" stands for asymptotically flat.
\end{Definition}
Condition i) makes sure that $\Psi$ is a solution of the ``Quantum 
Einstein Equations". Condition ii) is a possible way of defining 
asymptotic flatness $q_{ab}\to\delta_{ab},K_{ab}\to 0$ (although not very 
carefully, no fall-off and parity conditions were imposed \cite{13,13a} and
certainly this condition needs to be refined in future publications. For 
instance, one might imagine that the error of $\Psi(\hat{V}(R)\xi)/V_0(R)
-1$ is even smaller than $\ell_p^3/V_0(R)$ in the sense that it 
could depend on some negative power of the value
of the radius $r$ (with respect to $\delta_{ab}$) at the center of $R$).
Condition iii) makes contact with physics and allows us to identify certain 
states with elementary particles. In particular, in the present context of 
pure gravity we should be able to isolate the graviton (spin 2,
massless) states. Notice that in order to allow for non-trivial 
representations of the little subgroup of the Lorentz group we must
specify the appropriate Diffeomorphism group in the group averaging 
process in order to arrive at the diffeomorphism invariant states \cite{7}
which means, roughly speaking, that we include only those diffeomorphisms 
in the averaging process that approach identity at $\partial\Sigma$.
Finally, Condition iv) is the most imprecise one and attemts at defining
a possible quantum analogue for the dominant energy condition : recall 
that the classical energy momentum tensor $T_{\mu\nu}$ is said to satisfy 
the dominant energy condition if for every future directed timelike 
vector field $v^\mu$, the vector field $T^\mu_\nu v^\nu$ is zero or future 
directed and timelike. Now if $t^\mu=N n^\mu+N^\mu$ is the future 
directed timelike foliation vector field underlying the split 
$M=\Rl\times\Sigma$ and $n^\mu$ the normal vector field of $\Sigma$ 
then we may pick a frame such that $n^\mu T_{\mu\nu}t^\nu=N H_{matt}+N^a 
(V_{matt})_a$ is the only non-vanishing component of the vector 
$T_{\mu\nu}t^\nu$ and so we need to ask that it be non-negative, in 
particular, for $\vec{N}=0$, we ask that the matter Hamitonian densities 
be non-negative. This condition is, of course, incomplete since it is frame 
dependent and needs to be improved in the future. Notice that $(t^2 
g_{\mu\nu}t^\mu t^\nu)(x)=-t^2 N^2(x)+t^2 (q_{ab}N^a 
N^b)(x) \approx -N^2(x)t^2+ L(c(\vec{N},x,t))^2$ which motivates our 
definition of future directedness of $t^\mu$. 
We conclude with the remark that Condition iv) is certainly satisfied in the 
vaccum case that we are interested in in the present paper.\\
\\
One might think that a state satisfying all those conditions is rather 
hard to construct. Let us pause for a moment to argue that it is 
rather simple :

Consider for simplicity a $\Sigma$ with 
topology of $\Rl^3$ and distribute a countable number of vertices $v_n$ 
randomly into $\Sigma$ 
with an average next neighbour distance of a Planck length as measured by
$q^{(0)}_{ab}$. Make $v_n$ the only vertex of a 
graph $\gamma_n$ which is four-valent and non-co-planar (for instance,
$\gamma_n=\alpha_n\cap\beta_n$ where $\alpha_n,\beta_n$ are two kinks 
with vertex $v_n$). The graphs $\gamma_n$ are supposed to be  
contained in a box $B_n$ of $q^{(0)}$ volume $k\ell_p^3$ for some 
positive number $k$ and the $B_n$ are mutually 
non-inersecting. Consider normalized vectors $f_n$ which are finite linear 
combination of spin-network states
defined on $\gamma_n$ and which are eigenstates of the volume 
operator $\hat{V}(R)$ for any region $R$, all with the same eigenvalue 
$\lambda_n\ell_p^3=\lambda\ell_p^3>0$ if $v_n\in R$. Thus we have 
$\hat{V}(B_m)f_n=\lambda\ell_p^3\delta_{m,n}f_n$. Consider the 
infinite product state $\xi:=\prod_{n=1}^\infty f_n$ which is a regular 
(non-cylindrical) spin-network state on the infinite graph 
$\gamma=\cup_n \gamma_n$ and which is in fact
normalized, $||\xi||=1$ thanks to the disjointness of the graphs $\gamma_n$
because of which $||\xi||=\prod_n||f_n||$ due to the properties of the 
Ashtekar-Lewandowski measure. We now choose $k:=\lambda$
and find that for any macroscopic $R$, that is, any $R$ that contains 
many of the boxes $B_n$ it holds that 
$\hat{V}(R)\xi=V_0(R)[1+o(\ell_p^3/V_0(R)]\xi$. Now, since no state
which is cylindrical with respect to any of the 
graphs $\gamma_n$ can be in the image of the Hamiltonian constraint
\cite{8,9,10} it follows from its definition \cite{15} that the $\hat{\eta}$
operator reduces to group averaging with respect to the diffeomorphism group
because of which the group averaged diffeomorphism invariant state 
$\Psi=[\xi]=\hat{\eta}\xi$ is normalized as well with respect to the 
physical inner product \cite{15} $||\Psi||_{phys}^2=\Psi(\xi)=||\xi||^2=1$.
Thus indeed $\Psi([\hat{V}(R)-V_0(R)]\xi)=o(\ell_p^3/V_0(R))$ is satisfied.
It is clear that the construction can be repeated for the surface operator
as well because most of the intersections of the macroscopic surface 
$S$ with the $\gamma_n$ will not be in vertices of the $\gamma_n$ so 
that $\hat{V}(R),\hat{A}(S)$ can be simultanously diagonalized up to 
errors of order of $\ell_p^2/A_0(S)$. Thus, almost every $f_n$ can be 
chosen as a simultanous eigenvector of $\hat{V}(R),\hat{V}(S)$. Finally,
any macroscopic, for simplicity non-self-intersecting (any loop is 
product of these), loop $\alpha$ on our particular $\gamma$ is of the 
product form $\alpha=\circ^n 
\alpha_n^{k_n},\;\alpha_n\subset\gamma_n,\;k_n\in\{0,1\}$ where $\;k_n=0$ 
except for finitely many. The $SU(2)$ Mandelstam algebra is too 
complicated as to exhibit an explicit solution for $SU(2)$ so let us 
argue with an $U(1)$ substitute that the condition stated in definition
(\ref{def1}) is reasonable. For $U(1)$ we have  
$h_\alpha=\prod_{k_n=1}h_{\alpha_n}$. Now, if we 
choose for simplicity $\alpha_n=\gamma_n$ then 
$f_n=\sum_{k=-N}^N a_k h_{\alpha_n}^k$ where $\chi_k(g)=g^k$ is the 
character of the irreducible representation of $U(1)$ with weight $k$.
Since $T=\chi_k\chi_l=\chi_{k+l}$ the condition stated in the definition 
amounts to asking that (for $U(1)$) $1=\prod_{k_n=1}\sum_k 
|a_k|^2=\prod_{k_n=1}\sum_{k=-N+1}^N\bar{a}_k a_{k+1}$ up to some 
corrections. Indeed, if we could choose all $a_k$ to be equal 
($=1/\sqrt{2N+1}$) then the error would be  $1-\prod_{k_n=1} [1-1/(2N+1)]$ 
which is small provided that $\sum_{k_n=1}1=o(L(\alpha)/\ell_p)<<N$.
For instance we may choose $N=[L/\ell_p]^2$ where $L$ is the bound on the  
length of a macrosccopic loop that we wish to consider. 
This will suffice to motivate definition (\ref{def1}). Obviously one has 
to refine it but this seems impossible without the notion of coherent states
and will be left for future publications \cite{16}.\\
\\
We will now show that on an asymptotically flat state the ADM energy
operator as defined below is non-negative. We will not show that vanishing
energy corresponds to Minkowski space. Our definition of asymptotically
flat states as of yet is not restrictive or precise enough for that. 
As we will see, in order for the ADM operator to be densely defined we need 
that the following stronger condition :\\ 
\\
{\em ii)' A state $\Psi=\hat{\eta}\xi$ is said to be asymptotically flat
provided ii) of Definition \ref{def1} holds and in addition :\\
Let $\gamma$ be the (infinite) graph on which $\xi$ depends, $p$ a 
point in $\partial\Sigma\cap V(\gamma)$ and $B_t,\;t\in [0,1]$ any
homotopy of regions in $\Sigma$ such that $p\in B_t$ for each $t$ and
$B_0=\{p\}$. Then we require that for each such $p$ there exists 
$\epsilon>0$ such that $\xi$ is,
for each $0<t<\epsilon$, a finite linear combination of eigenstates 
with non-vanishing and $t-$independent eigenvalues of the volume 
operator $\hat{V}(B_t)$.}\\
\\
It is not clear that condition ii) implies ii)'
and if that should not be the case then we must add the requirement ii)'
stated as an additional restriction on ${\cal H}_{af}$ ! Notice, however, 
that ii)' is not unreasonable in the asymptotically flat context.\\
It will turn out in the course of the derivation that 
the positivity of energy theorem then holds if we impose one additional 
condition on the so already restricted space of states.

We write expression (\ref{22}) this time in the form (setting 
$S=\partial\Sigma$ right from the beginning)  
\ba \label{28}
E_{ADM}&=&\lim_{\epsilon\to 0}
-\frac{2}{\kappa}\int_{\partial\Sigma} dS_a E^a_i(x)
\frac{1}{\epsilon^3\sqrt{\det(q)}(x)}
\int_\Sigma d^3y \chi_\epsilon(x,y)(\partial_b E^b_i)(y)
\nonumber\\
&=& \lim_{\epsilon\to 0}-\frac{2}{\kappa}\int_{\partial\Sigma} dS_a E^a_i(x)
\frac{1}{V(x,\epsilon)}
\int_\Sigma d^3y \chi_\epsilon(x,y)(\partial_b E^b_i)(y)
\nonumber\\
&=& \lim_{\epsilon\to 0}
-\frac{2}{\kappa}\int_{\partial\Sigma} dS_a(x) E^a_i(x)
\frac{1}{V(x,\epsilon)} \int_\Sigma d^3y 
\chi_\epsilon(x,y)[G_i(y)-\epsilon_{ijk}A_b^j(y) E^b_k(y)] \nonumber\\
&=:& \lim_{\epsilon\to 0} E^\epsilon_{ADM}
\ea
where $G_i=\partial_a E^a_i+[A_a,E^a]_i$ is the Gauss law constraint.
Recall that $G_i=0$ only needs to hold in the interior of $\Sigma$ 
because the Lagrange multiplier $\Lambda^i$ of the Gauss constraint falls 
off like 
$1/r^2$ so that at $\partial\Sigma$ every function of $E^a_i,A_a^i$ is 
gauge invariant. More precisely we have the following : It is of 
interest by itself to derive the quantum Gauss law operator on a 
function of {\em smooth} connections cylindrical with respect to a graph
$\gamma$ :
\ba \label{28a}
&&\int_\Sigma d^3x \Lambda^i(x)\hat{G}_i(x)f_\gamma\nonumber\\
&=&-i\hbar\kappa\sum_{e\in E(\gamma)}\int d^3x\Lambda^i(x)\int_0^1 dt
\dot{e}^a(t)([\partial_{x^a}\delta_{ik}+\epsilon_{ijk}
A_a^j(x)]\delta(x,e(t))) (X^k_e(t)f_\gamma)\nonumber\\
&=&-i\hbar\kappa\sum_{e\in E(\gamma)}\int d^3x\Lambda^i(x) \int_0^1 dt
([-\frac{d}{dt}\delta_{ik}+\epsilon_{ijk}
\dot{e}^a(t)A_a^j(x)]\delta(x,e(t))) (X^k_e(t)f_\gamma)\nonumber\\
&=&-i\hbar\kappa\{\sum_{e\in E(\gamma)}\int d^3x\Lambda^i(x)
[-\delta(x,e(1))X_e^i(1)+\delta(x,e(0))X^i_e(0)\nonumber\\
&& + \int_0^1 dt\delta(x,e(t))(\dot{X}^i_e(t)+\epsilon_{ijk}
\dot{e}^a(t)A_a^j(x)X^k_e(t)]f_\gamma\}\nonumber\\
&=&-i\hbar\kappa\{\sum_{e\in E(\gamma)}
[-\Lambda^i(e(1))X_e^i(1)+\Lambda^i(e(0))X^i_e(0)\nonumber\\
&& + \int_0^1 dt\Lambda^i(e(t))(\dot{X}^i_e(t)+\epsilon_{ijk}
\dot{e}^a(t)A_a^j(e(t))X^k_e(t)]f_\gamma\} \;.
\ea
Here we have made use of $\partial_x\delta(x,y)=-\partial_y\delta(x,y)$ 
in the third step which holds on spaces of test functions of rapid decrease. 
But since for smooth connections $h_e(t,t+\delta t)^{\pm 1}=1\pm\dot{e}^a(t)
\delta t A_a(e(t))+o(\delta t^2)$ we have
\ba \label{28b}
&&X^i_e(t+\delta t)-X^i_e(t)
=\mbox{tr}(h_e(0,t+\delta t)\tau_i h_e(t+\delta t,1)
\partial_{h_e(0,1)})-X^i(t)\nonumber\\
&=&\mbox{tr}(h_e(0,t)h_e(t,t+\delta t)\tau_i h_e^{-1}(t,t+\delta t)
h_e(t,1)\partial_{h_e(0,1)})-X^i(t)\nonumber\\
&=&\delta 
t\dot{e}^a(t)\mbox{tr}(h_e(0,t)[A_a(e(t)),\tau_i]h_e(t,1)\partial_{h_e(0,1)})
+o(\delta t^2)\nonumber\\
&=&-\epsilon_{ijk}\delta 
t\dot{e}^a(t)A^j_a(e(t))\mbox{tr}(h_e(0,t)\tau_k h_e(t,1)\partial_{h_e(0,1)})
+o(\delta t^2)\nonumber\\
&=&-\epsilon_{ijk}\delta t\dot{e}^a(t)A^j_a(e(t)) X^k_e(t)
+o(\delta t^2)\;.
\ea
This shows that the $t-$ integral in (\ref{28a}) vanishes identically for 
smooth connections. Now $X^i_e(0)=X^i(h_e)=:X^i_e$ where $X^i(g)$ is the 
right invariant vector field at $g\in SU(2)$ and 
$X^i_e(1)=-X^i(h_e^{-1})$. Thus, when splitting each edge into 
two halves $e=e_1\circ e_2^{-1}$ 
where both $e_1,e_2$ are outgoing at the vertex different from their 
intersection point then $e(1)=e_2(0)$ and the right invariance of $X$
now implies $X(h_e)=X(h_{e_1}),\;-X(h_e^{-1})=X(h_{e_2})$. Summarizing we 
find that the Quantum Gauss Constraint is given by 
\be \label{28c}
\hat{G}(\Lambda)f_\gamma=-i\hbar\kappa\sum_{e\in E(\gamma)} 
\Lambda^i(e(0))X^i_e 
f_\gamma =-i\hbar\kappa\sum_{v\in V(\gamma)} \Lambda^i(v) X^i_v f_\gamma
\ee
where $-iX_v=-i\sum_{e(0)=v}X_e$ is the total ``internal" angular momentum 
operator.
Notice that (\ref{28c}) can be extended from smooth to distributional 
connections and that no assumption on asymptotic behaviour or smoothness 
of $\Lambda$ had to be made. 
The Quantum Gauss Constraint is obviously a self-adjoint 
operator on ${\cal H}$ and anomaly free : it is trivial to check that
$[\hat{G}(\Lambda),\hat{G}(\Lambda')]=\hat{G}([\Lambda,\Lambda'])$
precisely mirroring the classical constraint algebra. Moreover, the 
Quantum Gauss law constraint is 
identically satisfied as $\gamma$ tends to $\partial\Sigma$ because 
$\Lambda_{|\partial\Sigma}=0$, that is, there are no internal charges in 
general relativity \cite{0}. This allows quantum states of 
distributional connections to be non-gauge invariant at spatial infinity,
a fact that we are going to exploit in the sequel.

Let now the state $\xi\in {\cal H}_{af}$ be considered as a 
function $f_\gamma$ cylindrical with respect to a graph $\gamma$ which
is a finite subgraph of the graph on which $\xi$ depends and which 
inersects $\partial\Sigma$. Because $\xi\in {\cal H}_{af}$ we know that 
$f_\gamma$ is a {\em finite linear combination of eigenstates of the volume 
operator $\hat{V}(R)$ with non-zero eigenvalue for sufficiently small 
regions $R$ and such that $R\cap 
V(\gamma)_{|\partial\Sigma}\not=\emptyset$}. 
Consider first the volume integral in (\ref{28}). 
Setting $\Lambda(y)=\chi_\epsilon(x,y)$
there is an obvious quantization for the term proportional to $G_i$
in view of (\ref{28c}). However, for the remainder we have, again 
on functions of smooth connections only to begin with
\ba \label{29}
&& \epsilon_{ijk}\int_\Sigma d^3y \chi_\epsilon(x,y)
A_a^j(y)\hat{E}^a_k(y) f_\gamma \nonumber\\
&=& -i\hbar\kappa\epsilon_{ijk} \sum_{e\in E(\gamma)}\int_\Sigma d^3y 
\chi_\epsilon(x,y)A_a^j(y)
\int_0^1 dt \dot{e}^a(t) \delta(y,e(t))X^k_e(t) f_\gamma 
\nonumber\\
&=& -i\hbar\kappa \epsilon_{ijk}\sum_{e\in E(\gamma)}
\int_0^1 dt \dot{e}^a(t)A_a^j(e(t))\chi_\epsilon(x,e(t))
X^k_e(t) f_\gamma\nonumber\\ 
&=& 2i\hbar\kappa \sum_{e\in E(\gamma)}\lim_{n\to\infty}
\sum_{k=1}^n \chi_\epsilon(x,e(t_{k-1}))
\mbox{tr}(\tau_i[h_e(t_{k-1},t_k)-1])X^i_e(t_{k-1}) f_\gamma \;.
\ea
Now recall that $x\in\partial\Sigma$ and that
$\gamma\subset B(\partial\Sigma)$. Pick a particular edge $e$ in (\ref{29}).
Then for sufficiently small $\epsilon$ the corresponding term either 
vanishes or $e(0)=x$ and the term becomes 
$\mbox{tr}(\tau_i[h_e(0,t(\epsilon))-1])X^i_e f_\gamma+o(\epsilon^2)$
where $t(\epsilon)$ is the largest value of $t$ such that
$\chi_\epsilon(x,e(t))=1$. Now for a classical, smooth connection which 
approaches 
infinity as $1/r^2$ the term $h_e(0,t(\epsilon))-1$ is at most of order 
$\epsilon/r$ (the ``length" of $e(\epsilon)$ is at most of order $r$) and so 
vanishes even at finite $\epsilon$ because $x\to \partial\Sigma$. Therefore   
expression (\ref{29}) vanishes and the volume integral contribution of 
$\hat{E}_{ADM}$ becomes 
\be \label{30a}
\int_\Sigma d^3y \chi_\epsilon(x,y)
\partial_a\hat{E}^a_i(y) f_\gamma \nonumber\\
=-i\hbar\kappa \sum_{v\in V(\gamma)}
\chi_\epsilon(x,v)X^i_v f_\gamma
\ee
which one can extend to non-smooth connections.

We turn to the surface integral of (\ref{28}) and write 
$f_{\gamma,v}=X^i_v f_\gamma$ 
We have, ordering the $1/\hat{V}(x,\epsilon)$ to the left
\ba \label{30}
&&\int_{\partial\Sigma} \chi_\epsilon(x,v) 
dS_a\frac{1}{\hat{V}(x,\epsilon)}\hat{E}^a_i f_{\gamma,v}\nonumber\\
&=&-i\hbar\kappa\sum_{e\in E(\gamma)}\int_0^1 dt \dot{e}^a(t)
\int_{\partial\Sigma}dS_a(x)\delta(x,e(t))
\frac{\chi_\epsilon(x,v)}{\hat{V}(x,\epsilon)}X^i_e(t)f_{\gamma,v}
\nonumber\\
&=&-i\hbar\kappa\sum_{e\in E(\gamma),e(0)\in\partial\Sigma} 
\mbox{sgn}(\partial\Sigma,e) 
\frac{\chi_\epsilon(e(0),v)}{\hat{V}(e(0),\epsilon)}X^i_e f_{\gamma,v}
\nonumber\\
&=&i\hbar\kappa
\sum_{e\in E(\gamma),e(0)\in\partial\Sigma,e\not\subset\partial\Sigma}
\frac{\chi_\epsilon(e(0),v)}{\hat{V}(e(0),\epsilon)}X^i_e f_{\gamma,v}
\ea
where $\mbox{sgn}(S,e)$ is the sign of the intersection of $e$ with the 
surface $S$ (which is outward oriented) at
$e(0)$ which is thus $-1$ because all edges $e$ are outgoing from a vertex 
$e(0)$ and, because of $\gamma\subset B(\partial\Sigma)$, they are thus 
running away from 
$\partial\Sigma$ (there is no contribution from edges that run inside 
$\partial\Sigma$ because $\mbox{sgn}(\partial\Sigma,e)_{|e(t)}=0$ for all 
$t$ as was shown in \cite{ALVol}). Thus, only edges which run transversally
into $\partial\Sigma$ contribute to the sum in (\ref{30}).\\
We can take now the limit $\epsilon\to 0$. Notice that for small enough
$\epsilon$ we have $\chi_\epsilon(e(0),v)=1$ as $\epsilon\to 0$ provided
that $e(0)=v$ is a vertex of $\gamma$. Moreover, for small enough
$\epsilon$, $v$ is the only vertex of $\gamma$ in the $\epsilon-$box 
around $v$. Thus, we may replace $\hat{V}(v,\epsilon)$ by $\hat{V}_v$
interpreting the operator $1/\hat{V}_v$ by its spectral resolution. Now 
the only critical point is whether the two operators $X_e^i,X_v^i$
that are to the right of $1/\hat{V}_v$ will leave the crucial property of
$f_\gamma$ intact, namely, that 
$f_\gamma=\sum_{i=1}^N f_i$ with $\hat{V}_v f_i
=\lambda_i f_i,\;\lambda_i\not=0$. But this is 
easily seen to be the case provided we impose the following additional 
restriction on ${\cal H}_{af}$ :\\
\\
{\bf Tangle Property}\\
{\it An asymptotically flat state is a linear combination of diffeomorphism 
group averaged elements of the kinematical Hilbert space ${\cal H}_{af}$ 
each of which depends on an (infinite) graph $\gamma$ all of 
whose intersections with $\partial\Sigma$ are transversal, that is, there are
no edges of $\gamma$ which lie inside $\partial\Sigma$. Thus, any path
along $\gamma$ between 2 distinct points of $\gamma\cap\partial\Sigma$
is a generalized tangle \cite{BaTan}, that is, a piecewise analytic path 
intersecting $\partial\Sigma$ transversally and points of non-analyticity
are vertices of $\gamma$, equivalently, intersections of tangles.
The subspace of ${\cal H}_{af}$ having the tangle property will be called
${\cal H}_{tangle}$}.\\
\\
The tangle property is a rather natural assumption about states $\xi$ 
because curves of non-zero parameter length running in $\partial\Sigma$ 
themselves between vertices automatically have infinite length with 
respect to $\delta_{ab}$. This is not the case for curves between 
vertices which just approach $\partial\Sigma$ but otherwise lie 
in the interiour.\\
Under this additional assumption, we have in the limit $\epsilon\to 0$, 
putting (\ref{28c}), (\ref{30})
and the remaining pre-factor of $-2/\kappa$ from (\ref{28}) together
\be \label{31}
\hat{E}_{ADM}(S)f_\gamma=-2\hbar^2\kappa\sum_{v\in \partial\Sigma\cup 
V(\gamma)} \frac{1}{\hat{V}_v}X^i_v X^i_v f_\gamma\;.
\ee
But recognizing $-iX_v=\hat{J}_v$ as the total angular momentum of 
$f_\gamma$ at $v$ we finally find
\be \label{32}
\hat{E}_{ADM} f_\gamma=2\hbar^2\kappa\sum_{v\in\partial\Sigma\cup V(\gamma)}
\frac{1}{\hat{V}_v}\hat{J}^i_v \hat{J}^i_v f_\gamma\;.
\ee
Expression (\ref{32}) defines a self-consistent family of operators 
$\{\hat{E}_{ADM,\gamma}\}$ of operators which can be extended to infinite
graphs and thus to ${\cal H}_{tangle}$ provided that the number of 
punctures of $\gamma$ with $\partial\Sigma$ is finite. This defines 
the domain of $\hat{E}_{ADM}$.\\ 
A number of remarks are in order : 
\begin{itemize}
\item
The operator $\hat{J}_v^i$ is the infinitesimal generator of gauge 
transformations at $v$ and therefore commutes with $\hat{V}_v$ because the 
volume 
operator is gauge invariant proving that (\ref{32}) is densely defined on 
the restricted state space. If we would not require the tangle property
then while the volume integral in (\ref{28}) gives essentially $X_v$,
the surface integral does not, rather it gives 
$X'_v=X_v-\sum_{e(0)=v,e\subset\partial\Sigma}X_v$ which does not commute 
with
$\hat{V}_v$ and so may map $f_\gamma$ into a linear combination of states 
some of which may acquire zero volume and so (\ref{32}) would blow up.
Thus, the tangle property is {\em sufficient} for $\hat{E}_{ADM}$ to be 
a densely defined operator. Although we do not have a proof, it is almost
granted that $X'_v f_\gamma$ will contain zero volume eigenstates in its 
expansion and so the tangle property would also be {\em necessary} !
\item
As a striking bonus of the tangle property we easily see that (\ref{32})
indeed defines a positive-semidefinite self-adjoint operator on the 
Hilbert space ${\cal H}_{tangle}$ : First of all, since the 
volume operator and the $X^i_v$ are defined in terms of the operators 
$X^i_e$ it follows that the family of operators (\ref{32}) is 
consistently defined because the $X^i_e$ are. Next, since 
$[X^i_v,\hat{V}_v]=0$ we see that we may order (\ref{32}) symmetrically 
involving only terms of the form 
$\hat{J}^i_v\frac{1}{\hat{V}_v}\hat{J}_v^i$ and so (\ref{32}) defines a 
symmetric operator because it leaves the graph $\gamma$ invariant
\cite{2,3,4}. Finally, 
the Laplacian $-\Delta_v:=J_v^i J_v^i$ has non-negative
discrete spectrum $j_v(j_v+1)$ where $j_v$ is the spin of the contractor 
of the generalized (non-gauge-invariant) spin-network states into
which $f_\gamma$ can be decomposed.
Moreover, since $-\Delta_v$ and $\hat{V}_v$ commute,
they can be simultanously diagonalized. Thus, if $\lambda_v$ is 
the eigenvalue of the volume operator then the simultanous eigenstate
$f_\gamma$ is an eigenstate of the ADM energy with eigenvalue
$\sum_v\frac{j_v(j_v+1)}{\lambda_v}$. This provides an explicit 
diagonalization of $\hat{E}_{ADM}$ on ${\cal H}_{tangle}$ and demonstrates 
that it is a self-adjoint operator on ${\cal H}_{tangle}$.
Now, since states in ${\cal H}_{tangle}$ are not diffeomorphism averaged at 
spatial infinity the part of the graph which is responsible for the spectrum
of the ADM energy is untouched by the diffeomorphism group averaging. Next, 
the map $\hat{\eta}$ that 
makes a diffeomorphism invariant state a solution of the Hamiltonian 
constraint is a generalized projector at each vertex of the graph 
separately \cite{8,9,10,15}. Therefore, it is the identity map at those 
vertices of 
$\gamma$ that lie in $\partial\Sigma$ because for a gauge transformation 
$N\to\partial\Sigma$ the action of the constraint at those vertices is 
trivially zero. 
Thus the part of the state responsible for the spectrum 
and the adjointness relations of $\hat{E}_{ADM}$ 
is unchanged by the map $\hat{\eta}$ and are thus preserved when we go to 
the physical Hilbert space ${\cal H}_{phys}=\hat{\eta}{\cal H}_{tangle}$.
Moreover, the spectrum is entirely discrete and 
non-negative, thus we have proved {\em A Quantum Positivity of Energy 
Theorem} ! 
\item
Astonishingly, the proof of this theorem turned out to be surprisingly 
simple : the proof of the classical positivity theorem is much more 
complicated and uses the boundary conditions and the Einstein equations 
at various stages. Why did we not need (a quantum analogue of) these
assumptions ? The answer is that we actually did use them : We used them
in the definition of an asymptotically flat state, in particular, that
the volume of the state be non-vanishing and that it be a solution of the 
quantum Einstein equations. Since the energy functional is really given
by $E(N)=H(N)+E_{ADM}(N)$ where the Hamiltonian constraint only vanishes 
on a solution, likewise the ADM energy operator $\hat{E}_{ADM}$ only 
represents energy if we apply it to a solution $\Psi$ of $\hat{H}'(N)\Psi=0$.
Here we have written the dual $\hat{H}'(N)$ on $\Phi'$ of $\hat{H}(N)$ 
because, as stated above, solutions $\Psi$ lie actually in $\Phi'$
\cite{7,10}. Finally, we used a regularization of the operator consisting
1) in the restriction of the space of states to functions of classical 
connections (this means here that they are smooth and decay at 
infinity as $1/r^2$) and then 2) in the extension of the expression 
for the operator obtained to all of 
$\Ab$. In the derivation of the operator we made crucial use of the fact 
that a classical connection decays at infinity.\\
Now a subtle issue is the following : by definition a solution $\Psi$ 
satisfies $\hat{H}'(N)\Psi=0$ for any $N$ which vanishes at 
$\partial\Sigma$. But how about lapse functions $N$ that approach a 
constant value at $\partial\Sigma$ ? It is now not a consequence of 
the formalism any longer that $\hat{H}'(N)\Psi=0$ should hold, very much
like in the case of the Gauss constraint. The only guideline of what
to do is the classical theory and there it is indeed true that on 
classical solutions $(A_0,E_0)$ to the Einstein equations $E(N)$ just
equals $E_{ADM}(N)_{A=A_0,E=E_0}$ so that we will require that 
$\hat{H}'(N)\Psi=0$ even for asymptotically constant lapse.\\
Now, by definition the operator $\hat{\eta}$ acts like the identity 
operator at $\partial\Sigma$.
Therefore we conclude that any physical state has the property that it is 
annihilated by $\hat{H}'(N)$
even for $N=const.$ and moreover it is a linear combination of 
eigenstates with non-zero eigenvalue of the volume operator $\hat{V}_v$
for each vertex of the graph $\gamma$ on which $\xi$ depends 
(with $\hat{\eta}\xi=\Psi$) such that $v\in\partial\Sigma$. One might 
suspect that the number of states that satisfy this condition is rather 
tiny but the contrary is the case : the volume operator has the 
particular property that it does not change the graph or the labelling of 
that graph with spin quantum numbers. Now there exist an infinite number
of states which are annihilated by $\hat{H}'(N)$ just because the graph
or its labellings is of a particular type (see \cite{10} where such 
states were labelled by ``spin-webs", more precisely, ``sources" 
of spin-webs) and so one can construct eigenstates of the volume 
operator of such states while they are still annihilated by $\hat{H}'(N)$.
As an aside, this might shed some light on the issue of how to interprete 
those special solutions of the Hamiltonian constraint.\\
These remarks are sufficient to show that then 
$\hat{E}_{ADM}'(N)=\hat{E}_{ADM}(N)$. 
\item
The fact that the ADM energy operator is essentially diagonal on spin-network
states can be interpreted as saying that the spin-network representation 
is the {\em non-linear Fock representation} of quantum general relativity.
Namely, if we compare the spin that the edges of a spin-network carry 
with the occupation number of momentum modes of, say, the free Maxwell 
field then we may interprete this spin essentially as the occupation number
of a gravitational mode which is not labelled by a momentum but by an edge.
\item
Notice that each vertex of $\gamma$
which lies in $\partial\Sigma$ must be at least three-valent (not 
four-valent 
because the function does not need to be gauge invariant at infinity) in 
order that it can 
have non-vanishing volume. Also notice that the gravitational energy is
quantized in quanta of the Planck mass : the eigenvalues of the 
volume operator are multipla of $\ell_p^3$ so the eigenvalues of 
energy, according to (\ref{32}) are multipla of $\hbar^2\kappa/\ell_p^3=
\sqrt{\hbar/\kappa}=m_p$.
\item 
For a classical connection every function at $\partial\Sigma$ is gauge
invariant because it decays like $1/r^2$ which results in a holonomy
of order $exp(i/r)\to 1$, i.e., a trivial holonomy. In quantum theory 
this is lifted by the distributional nature of a connection, smooth 
connections are assigned zero volume by the Ashtekar Lewandowski measure 
in the space of distributional connections and are unimportant.
\item
Gauge invariant states
at $\partial\Sigma$ correspond to vanishing energy eigenvalue. Thus,
energy seems to sit at non-gauge invariant vertices.
We may interprete this observation as follows : The gravitational energy 
in a state labelled by an (infinite) graph $\gamma$ is concentrated 
at the vertices of $\gamma$, and energy flows from vertex to vertex along 
the edges of $\gamma$ in quantized packages labelled by the spin of those 
edges. Non-zero energy at a vertex corresponds to lack of gauge 
invariance at this vertex meaning that the spins that flow out or into
a vertex do not add up to zero. Now in the interiour of $\Sigma$ the 
Quantum Gauss Constraint requires that all spins add up to zero. We 
interprete this as the connection dynamics version of the 
geometrodynamics result that there 
is no energy location in general relativity in the interior of $\Sigma$,
gravitational energy can be gauged away locally, it is pure gauge. However,
while it can be pushed around at one's will, one cannot entirely delete
it, one can push it all the way to spatial infinity where it eventually
shows up in the form of a non-zero net spin flow at the vertices of $\gamma$
at $\partial\Sigma$. 
\item
The fact that (representations of) the $SU(2)$ gauge group of general 
relativity should play an essential role for the energy is very unexpected
from a geometrodynamics point of view where one never even talks about 
the $SU(2)$ gauge freedom. Even the classical ADM expression 
$\int_{\partial\Sigma} dS_a (q_{ab,b}-q_{bb,a})$ is manifestly gauge 
invariant, so how
did the $SU(2)$ gauge group enter the stage\footnote{This question actually 
arises already in connection with the spectrum of the geometrical
operators volume, area and length \cite{RSVol,JVol,RDP,TTVol,11}, however,
since irreducible representations also carry gauge invariant information
the fact that these operators have a spectrum which is determined by 
spin-quantum numbers of {\em gauge-invariant states} is maybe not that 
surprising. What {\em is} surprising for the energy operator is that
it is the spins of non-gauge invariant states which determines the 
spectrum.} ? The answer is the following :
Notice that while the classical ADM expression is manifestly gauge invariant,
it is not at all covariant : The derivatives that appear in
$\int_{\partial\Sigma} dS_a (q_{ab,b}-q_{bb,a})$ are not covariant 
derivatives (in fact if they were covariant derivatives then the energy would
vanish identically). This does not need to concern us because at 
$\partial\Sigma$ the diffeomorphisms that underly the diffeomorphism 
constraint must die off. Now as we showed in (\ref{7}), when writing 
$\int_{\partial\Sigma} dS_a (q_{ab,b}-q_{bb,a})$ in terms of $E^a_i$
this lack of three-diffeomorphism covariance, by means of the boundary 
conditions, gets translated into non-gauge-invariance while installing
covariance because in (\ref{7}) a correction term drops out 
due to the boundary conditions which explains why one and same 
function can be written in a manifestly gauge invariant or diffeomorphism 
covariant way but not both. Indeed, the expression
$-2\int_{\partial\Sigma} dS_a E^a_i \partial_b E^b_i/\sqrt{\det(q)}$ is 
manifestly 
three-diffeomorphism covariant but it fails to be gauge-invariant. 
This is not unexpected : after all the triad formulation reduces the local 
diffeomorphism gauge freedom to local rotation freedom.
In conclusion, the fact that states with non-zero energy
are not gauge invariant is in fact very natural.
\item
Another function for which $SU(2)$ gauge transformations and 
diffeomorphisms get
mixed up is the classical vector constraint $V_a=\mbox{tr}(F_{ab} E^b)$.
Strictly speaking this constraint function does not generate 
diffeomorphisms but only on gauge invariant functions.
\item
Notice that although
$-2\int_{\partial\Sigma} dS_a E^a_i \partial_b E^b_i/\sqrt{\det(q)}$ is not 
gauge invariant the quantum expression (\ref{32}) in fact is. The reason 
why that is 
possible lies in the structure of quantum theory : in the classical theory
we only have functions on phase space. In quantum theory those functions 
get translated into operators on a Hilbert space but values of those 
functions really correspond to expectation values. Thus non-gauge invariant
functions correspond to expectation values of either a gauge-invariant 
operator in a non-gauge-invariant state or vice versa. The quantization
(\ref{32}) picks the latter possibility.
\item 
In principle we have now solved the ``problem of time" : Since we 
have a true Hamiltonian we can introduce the {\em Schroedinger time parameter}
$t$ and our state vectors $\Psi\in\Phi'$, being distributions which are
invariant under asymptotically identity diffeomorphisms are 
supposed to satisfy the {\em non-stationary Schroedinger equation}
\be \label{32a}
\hat{E}_{ADM}(N=1)\Psi=-i\hbar\partial_t\Psi\;.
\ee
Notice that, as we showed above, the ADM energy operator is its own dual 
so that (\ref{32a}) makes sense. 
\item
In the next section the operator (\ref{32}) will be shown to commute with 
all quantum constraints of general relativity and it is therefore a
{\em Strong Quantum Dirac Observable} for Quantum Gravity in the strict 
sense of the word. This is not unexpected because it is built purely from
momentum operators.
\end{itemize}

\section{The Poincar\'e Algebra}

We now wish to define the rest of the quantum generators of the little 
group of the
asymptotic Poincar\'e group and check whether their algebra is anomalous 
or not \footnote{By little group of the Poincar\'e group we 
mean the group generated by the four-translations and the little subgroup of 
the connected component of the Lorentz Group. The latter, as is 
well-known, is the stabilator subgroup of the Lorentz group associated 
with a standard four vector $\bar{s}$. In the massive case $\bar{s}^2>0$
the standard vector is associated with the spin of the particle in the rest
frame and the covering group of the stabilator group is given by by $SU(2)$.
In the massless case the standard vector is associated with the helicity
of the particle (spin in momentum direction) and 
the covering group of the stabilator group is given by by $U(1)$, 
physically important representations being two-valued. Thus, the 
rotations at spatial infinity determine the unitary irreducible 
representation of the particle state in question.}. 
This is enough to construct particle states since the 
irreducible unitary representations of the little group induce a unique
unitary irreducible representation of the full Poincar\'e group. So far 
we did not construct an operator corresponding to a boost generator which
is more difficult to obtain than the ADM energy operator.

First of all we must clarify on which space to represent the Poincar\'e 
group, respectively its generators. To that end it is helpful to remember
how the classical Poincar\'e generators are realized as a subalgebra of 
the Poisson algebra \cite{13,13a}.\\ 
Let $H(N),V(\vec{N})$ be the Hamiltonian and diffeomorphism constraint 
functional respectively. Both functionals are integrals over $\Sigma$ of 
local 
densities and both converge and are functionally differentiable only
if the lapse and shift functions $N,\vec{N}$ vanish at $\partial\Sigma$.
In order to be able to describe the Poincar\'e group corresponding to
the asymptotically constant or even diverging functions ($x^a$ is a 
cartesian frame at spatial infinity)
$N=a+\chi_a x^a,N^a=a^a+\epsilon^{abc}\phi_b x^c$ where $(a,a^a)$ is 
a four translation, $\phi^a$ are rotation angles and $\chi^a$ are boost
parameters, one proceeds as follows : let $S$ be a bounded two-surface
that is topologically a sphere and let $B(S)$ be the (intersection of 
$\Sigma$ with the) closed ball such that $\partial B(S)=S$. For 
each $S$ one defines 
$E(N,S):=H(N,S)+E_{ADM}(N,S)+B(N,S),\;P(N,S):=V(\vec{N},S)+P_{ADM}(N,S)$
where the parameter $S$ means that volume integrals are restricted to
$B(S)$ only (a classical regularization of the divergent integrals) and the 
``counter-terms" $E_{ADM}(N,S),B(N,S),P_{ADM}(\vec{N},S)$ are the surface 
integrals defined in \cite{13} and correspond to ADM energy, boost and 
momentum. One can show that $\lim_{S\to\partial\Sigma} E(N,S), 
\lim_{S\to\partial\Sigma} P(\vec{N},S)$ exist. Moreover, for each finite $S$,
$E(N,S),P(\vec{N},S)$ are functionally differentiable so that it is 
meaningful to compute the Possion brackets
\ba \label{33}
\{E(M,S),E(M,S)\}=P(q^{ab}(MN_{,b}-M_{,b}N),S)\nonumber\\
\{E(M,S),P(\vec{N},S)\}=E({\cal L}_{\vec{N}}M,S)\nonumber\\
\{P(\vec{M},S),P(\vec{N},S)\}=P({\cal L}_{\vec{M}}\vec{N},S)\;.
\ea
The crucial point is that one computes the Poisson brackets a) at finite 
$S$ and b) on the full phase space and then takes the limit 
$S\to\partial\Sigma$ or restricts to the constraint surface of the phase 
space (where $H(N,S)=V(\vec{N},S)=0$). Notice that the numerical value 
of, say, $E(N,S)$ equals $H(N,S)$ for a gauge transformation for which $N\to 
0$ as $S\to\partial\Sigma$. On the other hand, on the constraint surface 
for a symmetry for which $N\not\to 0$ as $S\to \partial\Sigma$ it equals a 
time translation or a boost respectively. A similar remark holds for 
$P(\vec{N},S)$. One therefore interpretes (\ref{33}) as follows :
if $M,N$ are both pure gauge then the constraint algebra closes. If $M$
is a symmetry and $N$ pure gauge then energy (or boost generator) are gauge 
invariant.
If $M,N$ are both symmetry then time translations commute with each other,
time translations and boosts give a spatial translation and a boost with a 
boost gives a rotation, in other words the symmetry algebra closes.

In quantum theory we will therefore proceed as follows : \\
Recall \cite{8,9,10,15} that the Hamiltonian constraint $\hat{H}(N)$
(for asymptotically vanishing $N$) is only well-defined on the subspace
of $\Phi'$ corresponding to distributions on $\Phi$
which are invariant under diffeomorphisms that approach identity at 
$\partial\Sigma$. Thus we can expect the symmetry algebra to hold only on
such distributions as well. In fact, we will just choose $\Psi$ to be a 
solution to all constraints.\\ 
Next, in view of the fact that even the classical symmtry algebra 
only holds provided one first computes Poisson brackets at finite $S$ and 
then takes the limit, we will check the quantum algebra first 
by evaluating $\Psi$ on $\hat{E}(N_S) f_S$ 
for functions $f_S\in\Phi$ cylindrical with 
respect to a graph which lies in the interior of $B(S)$ (it may 
intersect $S$ in such a way that the volume operator does not vanish
at the intersection point for none of the eigenvectors into which
$f_S$ maybe decomposed) and lapse functions $N_S$ which grow at infinity like
symmetries but which are supported in $B(S)\cup S$ {\em including $S$}, and 
then to take the limit $S\to\partial\Sigma$ 
(the support fills all of $\Sigma$ as $S\to \partial\Sigma$ in this 
process). \\ 

We come to the definition of $\hat{E}(N),\hat{P}(\vec{N})$. 
First we treat the spatial Euclidean group.\\
The unitary representation of the diffeomorphism group defined by
$\hat{U}(\varphi) f_\gamma=f_{\varphi(\gamma)}$ which was for matters of 
solving the diffeomorphism constraint so far only defined for 
diffeomorphisms that approach asymptotically the identity, can easily 
be extended to three-diffeomorphisms which correpond to asymptotic spatial
translations or rotations. Instead of defining the generator 
$\hat{P}(\vec{N})$ though 
(which does not exist on $\cal H$ \cite{7}) we content ourselves with the 
exponentiated version $\hat{U}(\varphi(\vec{N}))$ where $\varphi(\vec{N})$
is the diffeomorphism generated by the six parameter shift vector field
$N^a=a^a+\epsilon_{abc} \phi^b x^c$ for some cartesian frame $x^a$ possibly 
corrected by an asymptotically vanishing vector field corresponding to a 
gauge transformation. It is trivial to check that 
\be \label{34}
\hat{U}(\varphi(\vec{N}))
\hat{U}(\varphi(\vec{N}'))
\hat{U}(\varphi(\vec{N}))^{-1}
\hat{U}(\varphi(\vec{N}'))^{-1}
=\hat{U}(\varphi({\cal L}_{\vec{N}}\vec{N}'))
\ee
where $\cal L$ denotes the Lie derivative 
so that there are no anomalies coming from the spatial Euclidean group.
This expression was derived by applying it to any function $f_S$ cylindrical 
with respect to a graph with support in $B(S)$. 

We now turn to the time translations. As already mentioned we will not 
consider boosts in this paper so that $\chi_a\equiv 0$ in 
the four parameter family of lapse functions $N=a+\chi_a x^a$
(modulo a correction which vanishes at $\partial\Sigma$).
Define the operator on ${\cal H}$
\be \label{35}
\hat{E}(N):=\hat{H}(N)
+\hat{E}_{ADM}(N)
\ee
where $\hat{H}(N)$ is the Lorentzian Hamiltonian constraint.
Notice that $\hat{E}(N)$ just as the Hamiltonian constraint in 
\cite{8,9,10,15} carries a certain prescription dependence 
which is removed by evaluating its dual on $\Phi_{Diff}$. We will not
repeat these details here and refrain from indicating this prescription
dependence in \ref{35}, however, the prescription dependence has 
consequences for the commutator algebra that we will dicuss below in 
great detail.

Let us verify the commutators between the time 
translations among themselves and between time taranslations and spatial 
translations and rotations. We have 
\ba \label{36}
&&\Psi([\hat{E}(M),\hat{E}(N)]f_\gamma)=
\Psi([\hat{H}(M),\hat{H}(N)]f_\gamma)
+\Psi([\hat{E}_{ADM}(M),\hat{E}_{ADM}(N)]f_\gamma)
\nonumber\\
&+&\Psi(\{[\hat{E}_{ADM}(M),\hat{H}(N)]+[\hat{H}(M),\hat{E}_{ADM}(N)]\}
f_\gamma)\;.
\ea
The first term vanishes for the same reason as in \cite{8,9,10,15} 
although one needs one additional argument : the 
Hamiltonian constraint does not act at vertices that it creates. Therefore,
it can be written as a double sum over vertices $v,v'$ of $\gamma$ alone and 
each of these terms is of the form 
$$
(M(v)N(v')-M(v')N(v))
\Psi([\hat{H}_{v',\gamma(v)}\hat{H}_{v,\gamma}-
\hat{H}_{v,\gamma(v')}\hat{H}_{v',\gamma}]f_\gamma)
$$
where the notation means that $\hat{H}_{v,\gamma}$ is a family of
consistently defined operators each of which  
acting on cylindrical functions which depend on the graph $\gamma$ and 
$\gamma(v)$ is a graph on which $\hat{H}_{v,\gamma}f_\gamma$ depends. 
This expression clearly is non-vanishing only if $v\not=v'$ but then 
it can be shown that the operators $\hat{H}_{v,..}$ and 
$\hat{H}_{v',..}$ actually commute. Now still this does not show that
the term above vanishes however, it can be shown that 
$\hat{H}_{v',\gamma(v)}\hat{H}_{v,\gamma}f_\gamma$ and 
$\hat{H}_{v,\gamma(v')}\hat{H}_{v',\gamma}f_\gamma$ are related by a 
diffeomorphism \cite{9}. Now in \cite{9} that was enough to show that the 
commutator vanishes because we were dealing there only with vertices 
which do not intersect $S$ as otherwise both lapse functions identically 
vanish for a pure gauge transformation. Thus the diffeomorphism that 
relates the two terms above could be chosen to have support inside 
$B(S)$ and $\Psi$ is invariant under such diffeomorphisms. In the present 
context that 
does not need to be true. However, the crucial point is now that by the 
tangle property all edges of $\gamma$ that intersect $S$ must intersect
$S$ transversally. Therefore the arcs that the Hamiltonian constraint 
attaches to $\gamma$ and whose position is the only thing by which the 
two above vectors differ {\em lie inside $B(S)$ and do not intersect $S$}. 
Therefore, again the two vectors are related by a diffeomorphism which 
has support inside $B(S)$, that is, they are related by a gauge 
transformation and therefore the commutator vanishes.

We turn to the second term in (\ref{36}). Now we obtain a double sum
over vertices of $\gamma$ which lie in $S$ and each term is 
of the form
$$
(M(v)N(v')-M(v')N(v))
\Psi([\hat{E}_{v',ADM},\hat{E}_{v,ADM}]f_\gamma)
$$
which is significantly simpler than before because $\hat{E}_{v,ADM}$
does not alter the graph. Notice that the commutator makes sense because
$\hat{E}_{ADM,v}$ leaves the span of non-zero volume eigenvectors invariant.
Now for $v\not=v'$ the commutator trivially vanishes, this time without 
employing diffeomorphism invariance of $\Psi$.

Finally the last term in (\ref{36}) is a double sum over vertices 
$v,v'$ of $\gamma$, where $v$ must lie in $S$, of the form
\be \label{37}
(M(v)N(v')-M(v')N(v))
\Psi([\hat{H}_{v',\gamma},\hat{E}_{v,ADM}]f_\gamma)\;.
\ee
The fact that $\hat{E}_{ADM}$ does not alter the graph was used to write
(\ref{37}) as a commutator without employing diffeomorphism 
invariance of $\Psi$. Now it may happen that, although $f_\gamma$ is in 
the domain of $\hat{E}_{v,ADM}$, that $\hat{H}_{v,\gamma}f_\gamma$ is not 
any longer in the domain and so (\ref{37}), for $v=v'$, is in danger of 
being 
a meaningless product of something that blows up times zero while that
cannot happen for $v\not=v'$. However, 
since $\Psi$ is a solution we conclude first of all that 
(\ref{37}) equals
\be \label{38}
-(M(v)N(v')-M(v')N(v))
[\hat{E}_{v,ADM}\Psi](\hat{H}_{v',\gamma} f_\gamma)
\ee
and since $\Psi$ is also in the domain of $\hat{E}_{ADM}$ both 
$\hat{E}_{v,ADM}\Psi$ and $\hat{H}_{v',\gamma} f_\gamma$
are well-defined elements of $\Phi'$ and $\Phi$ respectively we conclude 
that in case $v=v'$ (\ref{37}) indeed vanishes. On the other hand,
the same argument as before shows that the commutator trivially vanishes for 
$v\not=v'$.

Let us now check the commutator between time translations and spatial
translations and rotations $\varphi$. We have
\ba \label{39}
&&\Psi([\hat{U}(\varphi)^{-1}\hat{E}(N)\hat{U}(\varphi)-\hat{E}(N)]f_\gamma)
\nonumber\\
&=& \sum_{v\in V(\gamma)}[N(\varphi(v))
\Psi(\hat{U}(\varphi^{-1})\hat{H}_{\varphi(v),\varphi(\gamma)}
f_{\varphi(\gamma)})
-N(v)\Psi(\hat{H}_{v,\gamma}f_\gamma)]
\nonumber\\
&+& 
\sum_{v\in 
V(\gamma)\cap 
S}[N(\varphi(v))\Psi(\hat{U}(\varphi^{-1})\hat{E}_{ADM,\varphi(v)} 
f_{\varphi(\gamma)})-N(v)\Psi(\hat{H}_{ADM,v}f_\gamma)]\;.
\ea
Since $\hat{E}_{ADM}$ does not change the graph on which a function depends
we have identically 
$\hat{U}(\varphi^{-1})\hat{E}_{ADM,\varphi(v)} f_{\varphi(\gamma)}
=\hat{E}_{ADM,v} f_\gamma$.\\  
Now, as explained in more detail in \cite{9}, the operator $\hat{H}(N)$
depends on a certain prescription of how to attach loops to graphs. Since 
in the interiour of $B(S)$ there is no background metric
available, this prescription can only be topological in nature and therefore
graphs differing
by a diffeormorphism $\varphi$ are assigned graphs by $\hat{H}(N)$ which
are diffeomorphic by a diffeomorphism $\varphi'$ which may not coincide with
$\varphi$. That is, in the interiour of $B(S)$, $\hat{H}(N)$ is 
only covariant up to a diffeomorphism. On the other hand,
since one has the fixed background metric $\delta_{ab}$ at $S$ one can 
make $\hat{H}(N)$ precisely covariant at $S$, that is, the prescription 
satisfies $\varphi_{|S}=\varphi'_{|S}$.
Therefore, with this sense of covariance of $\hat{H}(N)$ 
it is true that 
$\hat{U}(\varphi^{-1})\hat{H}_{\varphi(v),\varphi(\gamma)}f_{\varphi(\gamma)}$
and $\hat{H}_{v,\gamma}f_\gamma$ differ at most by a diffeomorphism 
with support in the interiour of $B(S)$.\\ 
In conclusion we obtain
$$
\Psi([\hat{E}(N),\hat{U}(\varphi)]f_\gamma)=
\Psi(\hat{E}(\varphi^\star N-N)f_\gamma)
$$ 
which is what we were looking for.

We conclude that the little algebra of the Poincar\'e algebra is faithfully 
implemented.\\ \\
\\
\\
{\large Acknowledgements}\\
\\
This research project was supported in part by DOE-Grant
DE-FG02-94ER25228 to Harvard University.

\end{document}